# Vortex induced motion of the cylinder is suppressed by wearing a closed membrane

Yakun Zhao[1], Xinliang Tian[1,2,*]

[1]State Key Laboratory of Ocean Engineering, Shanghai Jiao Tong University, Shanghai 200240, China

[2]SJTU-Sanya Yazhou Bay Institute of Deep Sea Science and Technology, Sanya 572024, China

[*]Email address for correspondence: tianxinliang@sjtu.edu.cn

**Abstract**

Vortex-induced motion (VIM) of cylindrical structures in fluid is a common occurrence in nature and engineering. Over the years, numerous techniques for suppressing VIM have been developed. In this study, we propose a novel method for VIM suppression by applying a closed membrane to the cylinder. A demonstration experiment was conducted in a circulation water channel, varying the reduced velocity ($U^*$) from 2 to 11. The results demonstrate that the closed membrane effectively suppresses VIM, particularly in the lock-in region of $5 \leqslant U^* \leqslant 9$, where the motion amplitude of the cylinder is reduced by up to 85%. Additionally, the in-line mean drag force is also reduced when compared to that of a bare cylinder. It is important to note that this study is preliminary. However, the evidence obtained thus far is highly significant.

**Introduction**

For the flow past cylindrical structures, alternating vortex shedding occurs and fluctuating lift force acting on the body is generated. This may cause large resonance vibrations and motions phenomena, e.g., vortex-induced vibration (VIV) for slender risers [1] and vortex-induced motion (VIM) for floating offshore platforms [2], which might bring severe damages to structures and threats to operations. Therefore, numerous approaches have been developed to suppress VIV or VIM. The rigid appendage is considered as one of the most effective passive flow control methods to suppress vibrations. The most widely used appendages are helical strakes which have been reported to bring obvious vibration suppression [3], where the response amplitude can be reduced up to 80%. However, the helical strakes increase the nominal diameter of the structures and increase the fluid drag forces which causes larger displacement in the in-line direction. Another widely proposed appendages are split plates installed at the rear side of the cylinder. Obviously, the performance of the method is dependent on the direction of the incoming flow.

Recently, Gao *et al*. (2020), for the first time, proposed that the attachment of a closed membrane to the bluff body brings a considerable drag reduction up to 10% [4]. The closed membrane self-adapts to a streamlined shape under the interaction of internal and outer fluid flows, which may bring flow control benefits, i.e., not only in drag reduction, but also in vibration suppression, noise reduction and etc [5]. Motivated by this, here we test the performances of the closed membrane on the VIM response of a circular cylinder.

**Experimental method**

The experiments were conducted in the circulation water channel of the State Key Laboratory of Ocean Engineering (SKLOE) at Shanghai Jiao Tong University. The test section of the water channel has dimensions of 3.0 m in width, 1.6 m in depth and 8.0 m in length. The flow velocity ranges from 0 to 3 m/s.

Schematic diagram of the experimental setup is shown in Figures 1(a) and 1(b). The circular cylinder model used in the experiments is made of Plexiglas with a diameter of $D = 0.2$ m (see Figure 1c). The circular cylinder has a total height ($H$) of 0.7 m and its draft ($T$) is 0.5 m. The cylinder is vertically placed and moored with four identical mooring lines horizontally spread, as shown in Figure 1(a). The mooring lines is connected to the fairlead holes of the cylinder which is 0.01 m above the mean water level. The membrane model is made of polyethylene. Referring to the results by Gao *et al.* (2020), the circumference ($C$) of the flexible membrane is chosen as $C = 3.9\ D$. The length of the membrane is the same as the draft of the cylinder, 0.5 m. One end cap, also made of polyethylene, is used to close the membrane at the bottom of the cylinder. Then, the membrane completely envelops the underwater part of the circular cylinder, and the water (fluid) fills the space between the membrane and the cylinder, as shown in Figure 1(d). The reduced velocity, defined by $U^*=UT/D$, is investigated over the range of $2 \leq U^* \leq 11$, in which $U$ is the free-stream flow velocity. The corresponding Reynolds number ranges $20000 \leq Re \leq 110000$.

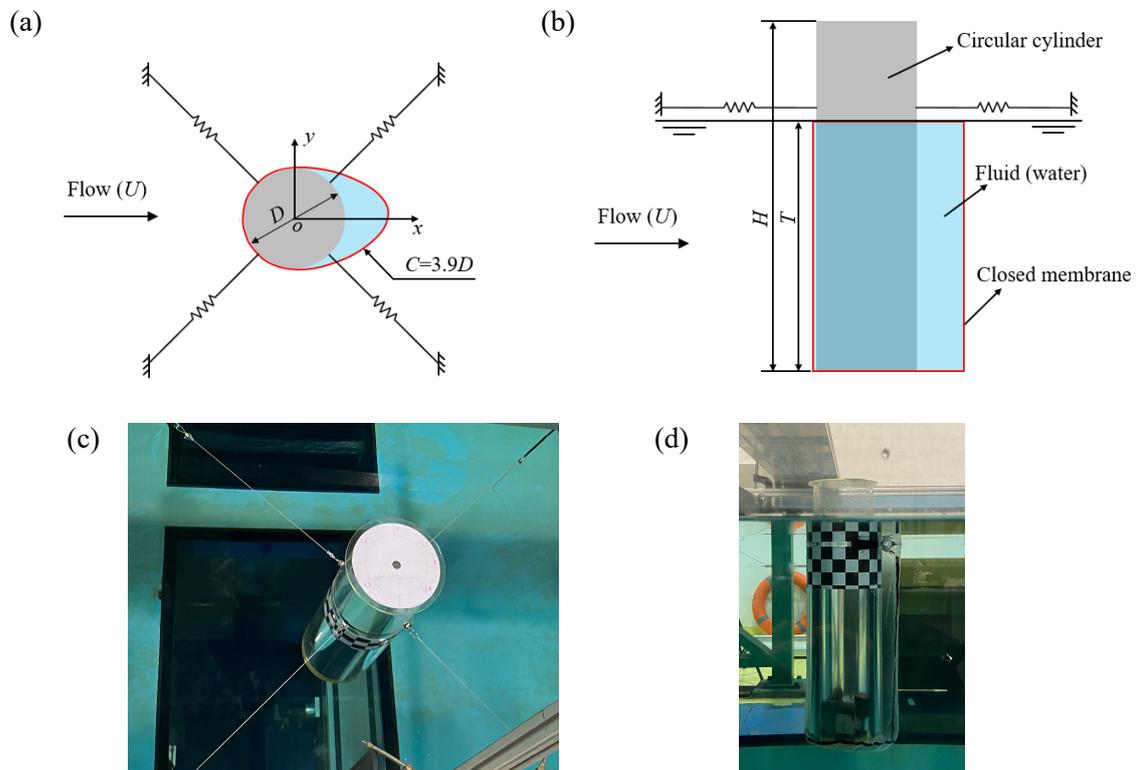

Figure 1. Schematic diagram of the experimental setup and definition of the coordinate system (a, b), and photograph of the experimental setup (c, d).

## Results and discussion

The motion trajectories of the bare cylinder and the cylinder with the closed membrane are shown in Figures 2(a) and 2(b), respectively. When $U^*$ is small, i.e., $U^*=3$, the bare cylinder and the cylinder with the closed membrane have very similar patterns of motion with small oscillations around their equilibrium position. As $U^*$ increases, the range of the motion of both the bare cylinder and the cylinder with the closed membrane becomes larger. For the bare cylinder, at $U^* > 3$, its motion amplitude increases sharply, and the shape of the motion trajectory is typically bow-shaped. When $U^*$ continues to increase to 9, the motion of the bare cylinder becomes chaotic, with both violent lateral and longitudinal motions. When $U^*$ exceeds 11, the trajectory of the cylinder returns to be regular, and the response of both lateral and longitudinal motions decreases dramatically. However, the motion observations of the cylinder with the closed membrane varying $U^*$ show significant differences with those of the bare cylinder. As $U^*$ increases, the motion of the cylinder with the closed membrane increases more gently. At $U^* >3$, the motion trajectory is typically a regular bow-shape at all $U^*$. Both lateral and longitudinal motion responses are significantly suppressed.

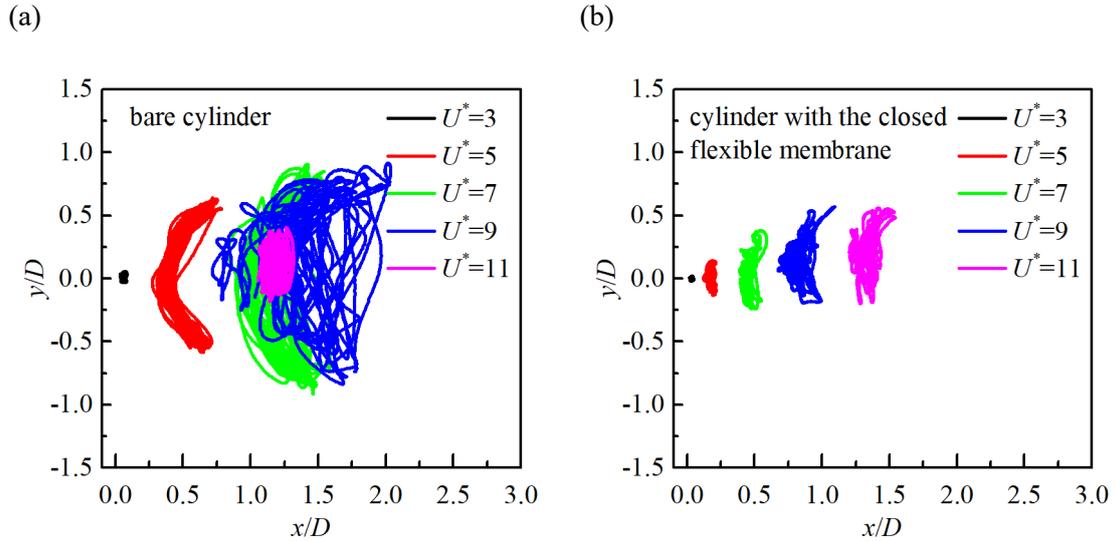

Figure 2. Motion trajectories in XY plane for (a) the bare cylinder and (b) the cylinder with the closed membrane.

Figure 3 shows the averaged amplitude ($A_y/D$) of the transverse motion at various $U^*$. For the bare cylinder, as $U^*$ increases, a clear lock-in range at $5 \leq U^* \leq 9$ is observed for the bare cylinder. Maximum value of the nominal transverse amplitude reaches about $A/D= 0.79$ at $U^* =7$. For the cylinder with the closed membrane, in general, the transverse motion slowly increases with the growth of $U^*$. The cylinder with the closed membrane has a maximum value of the nominal transverse amplitude of $A_y/D = 0.21$ at $U^* =9$. Compared to the bare one, in the lock-in range of $5 \leq U^* \leq 9$, the cylinder with the closed membrane decreases significantly, with a maximum reduction of approximately 85%. In the $U^*$ range far from the lock-in region, transverse motion amplitude of the bare cylinder and the cylinder with the closed membrane are very close.

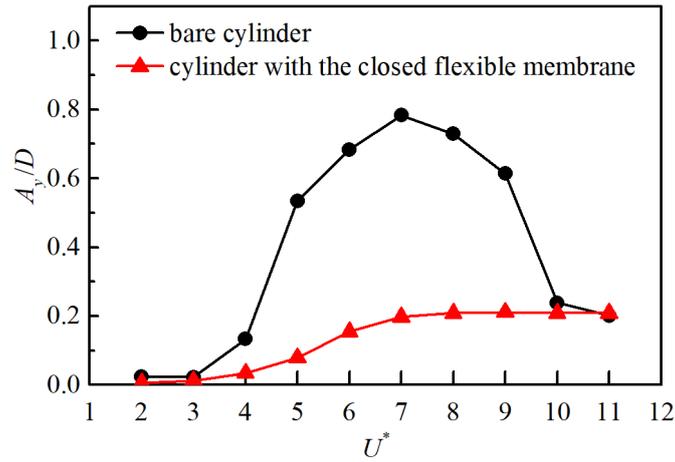

Figure 3. Response of the transverse motion amplitude for the bare cylinder and the cylinder with the closed membrane.

Results of the mean in-line displacement for the bare cylinder and the cylinder with the closed membrane are compared in Figure 4. At $5 \leq U^* \leq 9$, the mean in-line displacement of the bare cylinder has larger values, approximately three times, than that of the cylinder with the closed membrane. This indicates that the cylinder obtains a considerable drag reduction from the closed membrane.

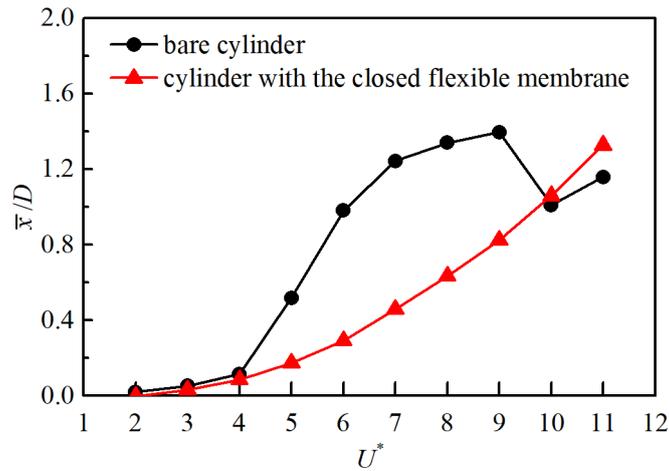

Figure 4. Mean in-line displacement for the bare cylinder and the cylinder with the closed membrane.

**Conclusion**

We propose a novel method for suppressing vortex-induced motion (VIM) by adding a closed membrane to a circular cylinder in flow. A preliminary experiment was conducted in a water circulation tunnel, considering a reduced velocity range ($U^*$) from 2 to 11. The bare cylinder exhibited noticeable VIM within the lock-in range of $5 \leq U^* \leq 9$, with a maximum transverse motion amplitude ($A_y/D$) of 0.79. In contrast, the presence of the closed membrane significantly suppressed the transverse motion of the cylinder, resulting in a maximum reduction of 85%. Additionally, the in-line drag force is also reduced when compared to a bare cylinder.

It is important to note that the results presented in this study are preliminary. However, the proposed VIM suppression method has demonstrated excellent performance. In comparison to traditional VIM suppression methods, this new approach not only suppresses transverse motion but also reduces the in-line drag force. Furthermore, this method is expected to outperform split-plate type methods in terms of its ability to handle flow from various directions. The authors believe that this new method, utilizing flexible materials, holds great potential for VIM suppression applications and deserves further investigation.

Acknowledgement
Xinliang Tian would like to thank Prof. Shijun Liao for his continuous support to the team working on this fundamental research. The assistance from Mr. Yi Dai in experiments is also appreciated.

References
[1] Blevins, R. D. & Saunders, H. 1997 Flow-induced vibration. Journal of Mechanical Design, 101(1):6.
[2] Waals, O.J., Phadke, A.C., Bultema, S. 2007 Flow induced motions on multi column floaters. In: ASME 2007 26th International Conference on Offshore Mechanics and Arctic Engineering. American Society of Mechanical Engineers, 669–678.
[3] Zdravkovich, M. M. 1981 Review and classification of various aerodynamic and hydrodynamic means for suppressing vortex shedding. Journal of Wind Engineering and Industrial Aerodynamics, 7(2):145-189.
[4] Gao, S., Pan, S., Wang, H., Tian, X. 2020 Shape deformation and drag variation of a coupled rigid-flexible system in a flowing soap film. Physical Review Letters, 125 (3):034502.
[5] Tian, X. Flexible coating reduces drag. 2021 Journal of Shanghai Jiao Tong University, 55:2 (in Chinese)